\documentclass[5p,a4paper,twocolumn,preprint,sort&compress,number]{elsarticle}

\usepackage[utf8]{inputenc}
\usepackage[T1]{fontenc}
\usepackage[english]{babel}
\usepackage{graphicx}
\usepackage{xcolor}
\usepackage{newtxtext}
\usepackage{mathtools}
\usepackage[vvarbb,subscriptcorrection]{newtxmath}
\usepackage{bm}
\usepackage[cal=cm,frak=euler]{mathalfa}
\usepackage{siunitx}
\usepackage[final]{microtype}
\usepackage{hyperref}

\newcommand{\mytitle}{Finite-volume effects in baryon number fluctuations around the QCD critical endpoint}

\DeclareSIUnit{\MeV}{\mega\electronvolt}
\DeclareSIUnit{\GeV}{\giga\electronvolt}
\DeclareSIUnit{\fm}{\femto\meter}
\DeclareMathOperator{\Tr}{Tr}
\DeclareMathOperator{\arsinh}{arsinh}

\renewcommand*{\vec}[1]{\bm{#1}}
\newcommand{\+}{\hspace*{.08335em}}

\newcommand{\dd}{\mathrm{d}}
\newcommand{\ii}{\mathrm{i}}
\newcommand{\bbR}{\mathbb{R}}
\newcommand{\bbZ}{\mathbb{Z}}
\newcommand{\bbN}{\mathbb{N}}
\newcommand{\calZ}{\mathcal{Z}}
\newcommand{\muCEP}{\mu_{\textup{B}}^{\textup{CEP}}}
\newcommand{\TCEP}{T_{\textup{CEP}}}

\newcommand{\Nc}{N_{\textup{c}}}
\newcommand{\Ksum}{K_{f}^{\+\textup{sum}}}
\newcommand{\Kint}{K_{f}^{\+\textup{int}}}
\newcommand{\OO}{\operatorname{O}}
\newcommand{\gs}{g}
\newcommand{\massu}{m_{\textup{u}}}
\newcommand{\massd}{m_{\textup{d}}}
\newcommand{\masss}{m_{\textup{s}}}
\newcommand{\muu}{\mu_{\textup{u}}}
\newcommand{\mud}{\mu_{\textup{d}}}
\newcommand{\mus}{\mu_{\textup{s}}}
\newcommand{\muB}{\mu_{\textup{B}}}
\newcommand{\muS}{\mu_{\textup{S}}}
\newcommand{\muQ}{\mu_{\textup{Q}}}
\newcommand{\chinX}[2]{\chi_{#1}^{\textup{#2}}}

\graphicspath{{./Graphics/}}

\hypersetup{%
	pdftitle = {\mytitle},
	pdfauthor = {Julian Bernhardt, Christian S. Fischer, Philipp Isserstedt},
	colorlinks = true,
	bookmarksopen = true
}

\journal{Physics Letters B}
\date{}

\definecolor{dgreen}{rgb}{0.1,0.5,0.1}
\definecolor{lblue}{rgb}{0.2,0.35,1}
\definecolor{webred}{rgb}{0.75,0,0}

\begin{document}

\begin{frontmatter}

\title{\mytitle}

\author[jlu,hfhf]{Julian Bernhardt}
\ead{julian.bernhardt@physik.uni-giessen.de}

\author[jlu,hfhf]{Christian S.~Fischer}
\ead{christian.fischer@theo.physik.uni-giessen.de}

\author[jlu,hfhf]{Philipp Isserstedt\fnref{fce}}
\ead{philipp.isserstedt@frankfurtconsultingengineers.de}

\address[jlu]{%
	Institut f\"{u}r Theoretische Physik,
	Justus-Liebig-Universit\"{a}t Gie\ss{}en,
	35392 Gie\ss{}en,
	Germany
}

\address[hfhf]{%
	Helmholtz Forschungsakademie Hessen f\"{u}r FAIR (HFHF),
	GSI Helmholtzzentrum f\"{u}r Schwerionenforschung,
	Campus Gie\ss{}en,
	35392 Gie\ss{}en,
	Germany
}

\fntext[fce]{%
	Present address:
	FCE Frankfurt Consulting Engineers GmbH,
	Bessie-Coleman-Str.~7,
	60549 Frankfurt am Main,
	Germany.
}

\begin{abstract}
We present results for the volume dependence of baryon number fluctuations
in the vicinity of the (conjectured) critical endpoint of QCD. They are
extracted from the nonperturbative quark propagator that is obtained as a
solution to a set of truncated Dyson--Schwinger equations of ($2 + 1$)-flavor
QCD in Landau gauge, which takes the backcoupling of quarks onto the
Yang--Mills sector explicitly into account. This well-studied system predicts
a critical endpoint at moderate temperatures and rather large chemical
potential. We investigate this system at small and intermediate finite,
three-dimensional, cubic volumes and study the resulting impact on baryon
number fluctuations and ratios thereof up to fourth order in the region of the
critical endpoint. Due to the limitations of our truncation, the results
are quantitatively meaningful only outside the critical scaling region of the
endpoint. We find that the fluctuations are visibly affected by the finite
volume, particularly for antiperiodic boundary conditions, whereas their ratios
are practically invariant.
\end{abstract}


\end{frontmatter}

\section{\label{introduction}%
	Introduction
}

Fluctuations of the conserved quantities baryon number, electromagnetic charge,
and strangeness are important quantities in our quest to locate a potential
critical endpoint (CEP) in the phase diagram of strong-interaction matter
\cite{Luo:2017faz}. It is one of the main goals of the experimental programs at
the Relativistic Heavy-Ion Collider at Brookhaven National Laboratory to find
such an endpoint in a dedicated beam energy scan \cite{Bzdak:2019pkr} that
probes a large baryon-chemical-potential area of the phase diagram, which is
not accessible at the Large Hadron Collider. The search for the CEP is
also one of the main motivations for the future Compressed Baryonic Matter
experiment \cite{Friman:2011zz} at the Facility of Antiproton and Ion Research
in Darmstadt and the Nuclotron-based Ion Collider Facility at the Joint
Institute for Nuclear Research in Dubna.

While at small chemical potentials an analytic crossover from the
low-temperature hadronic phase to a high-temperature quark-gluon plasma phase
has been firmly established, see Refs.~\cite{HotQCD:2019xnw,Bazavov:2020bjn,%
Borsanyi:2020fev,Guenther:2020jwe,Borsanyi:2021sxv} and references therein, the
location where the crossover turns into a first-order phase transition---marked
by a second-order CEP---is up to now not unambiguously determined. Elaborate
calculations within the functional approach to QCD using sets of coupled
Dyson--Schwinger or functional renormalization group (FRG) equations
\cite{Fischer:2014ata,Isserstedt:2019pgx,Fu:2019hdw,Gao:2020qsj,Gao:2020fbl,%
Gunkel:2021oya} place the endpoint into the region
$(\muCEP,\, \TCEP) = (\numrange{495}{654}, \numrange{108}{119}) \, \si{\MeV}$,
i.e., within a narrow temperature range with only moderate spread in chemical
potential (see Fig.~\ref{fig:phase_diagrams} below).

In experiments, signals for the CEP but also for the crossover and
the first-order phase transition at large baryon chemical potential are
expected to appear in ratios of event-by-event fluctuations, i.e., cumulants
and related quantities. These may be particularly prominent in higher orders
\cite{Stephanov:1998dy,Stephanov:1999zu,Asakawa:2009aj,Stephanov:2011pb,%
Friman:2011pf,Skokov:2012ds,Bzdak:2016sxg}. Ratios of cumulants have the
advantage that the explicit volume dependence of the fluctuations drops out.
However, as pointed out and investigated in Refs.~\cite{Skokov:2012ds,%
Almasi:2016zqf}, implicit volume dependences may remain. Whether these have
to be taken into account when comparing theoretical calculations with
experimental results is under debate. In any case, on systematic grounds it
is interesting to study these implicit dependences.

This is the topic of the present work. We study the explicit and implicit volume
effects in baryon number fluctuations and ratios thereof in continuum QCD using
the functional framework of Dyson--Schwinger equations (DSEs) at nonzero
temperature and chemical potential. As a first step, we use the
truncation scheme of Refs.~\cite{Isserstedt:2019pgx,Bernhardt:2021iql} and
investigate finite-volume effects for a range of cubic spatial volumes in the
CEP region of the $\muB$-$T$ plane. Since this scheme does not take the
effects of long-range degrees of freedom (e.g., those with the quantum numbers
of the sigma meson) into account, we do not expect to be able to study the full
volume dependence inside the critical scaling region. However, we believe that
the scheme at hand provides a meaningful starting point for the study of
fluctuations outside the critical scaling region. Furthermore, the truncation
can be improved systematically in the future along the lines of the (more
involved) scheme explored in Ref.~\cite{Gunkel:2021oya}.

We proceed as follows. First, our framework is detailed in
Sec.~\ref{framework}. Then, we summarize the implementation of a finite
spatial volume and the employed truncation and provide the reader with
information how we determine fluctuations. Then, in Sec.~\ref{results},
we discuss our results, and we conclude in Sec.~\ref{summary}.

\begin{figure*}[t]
	\centering%
	\includegraphics[scale=1.0]{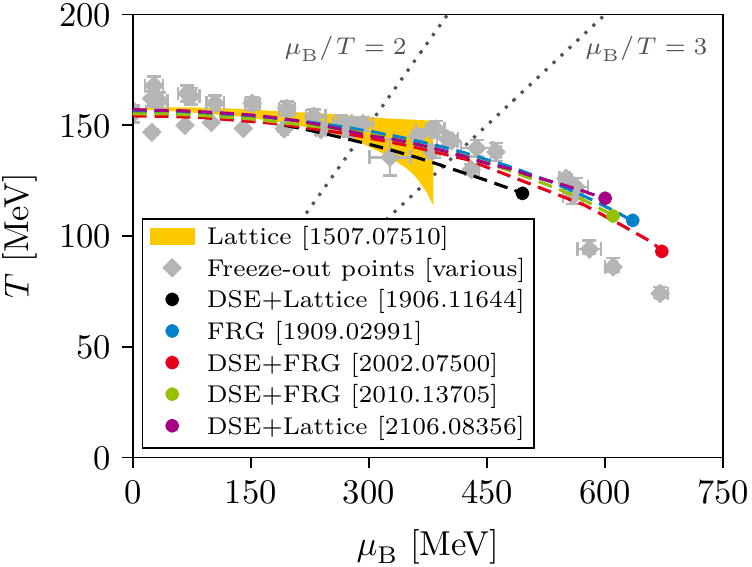}%
	\\[1.5em]%
	\includegraphics[scale=1.0]{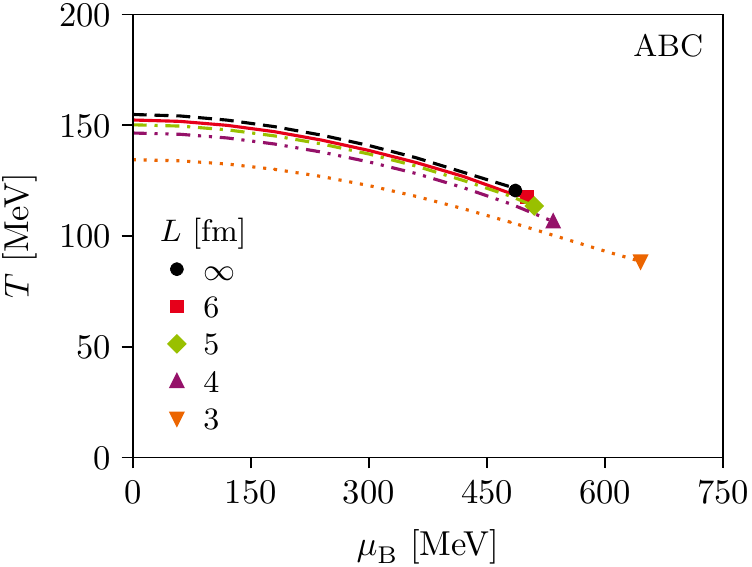}%
	\hspace*{1em}%
	\includegraphics[scale=1.0]{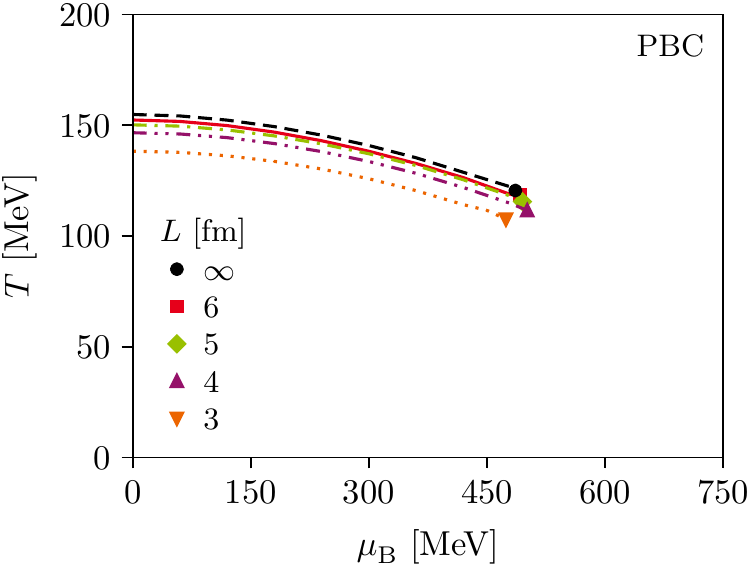}%
	\caption{\label{fig:phase_diagrams}%
		Upper diagram:
		various results for the analytic crossover and the CEP from
		state-of-the-art functional calculations \cite{Isserstedt:2019pgx,%
		Fu:2019hdw,Gao:2020qsj,Gao:2020fbl,Gunkel:2021oya} in comparison to
		an extrapolation from lattice QCD \cite{Bellwied:2015rza} and
		experimental freeze-out points \cite{Alba:2014eba,Vovchenko:2015idt,%
		Becattini:2016xct,Andronic:2016nof,STAR:2017sal,Andronic:2017pug}.
		Lower diagrams:
		volume dependence of the CEP of Ref.~\cite{Isserstedt:2019pgx} as
		determined in Ref.~\cite{Bernhardt:2021iql} using antiperiodic (ABC)
		and periodic (PBC) spatial boundary conditions. We use a cubic volume
		$V = L^3$, where $L$ denotes the edge length.
	}%
\end{figure*}

\section{\label{framework}%
	Framework
}

\subsection{\label{framework:finite_volume}%
	Finite-volume generalities
}

Introducing a finite, uniform, three-dimensional cubic volume with edge length
$L$ amounts to the restriction of all appearing three-dimensional
configuration-space integrals to the domain $[0, L]^{3} \subset \bbR^{3}$.
Imposing either periodic or antiperiodic boundary conditions (PBC or ABC,
respectively) in the spatial directions, each momentum-space integral is
therefore replaced by a sum over discrete spatial modes according to
\begin{equation}
	\label{eq:int_to_sum}
	\int \frac{\dd^{3} q}{(2 \pi)^{3}} \, (\ldots)(\vec{q})
	\to
	\frac{1}{L^{3}} \sum_{\vec{\xi} \in \bbZ^{3}}
	(\ldots)(\vec{q}_{\vec{\xi}})
\end{equation}
with $\vec{q}_{\vec{\xi}} = [\varpi_{\xi_{1}}, \varpi_{\xi_{2}},
\varpi_{\xi_{3}}]^{\top}$ and $\vec{\xi} \in \bbZ^{3}$, where the
spatial Matsubara modes are given by ($i = 1, 2, 3$)
\begin{equation}
	\label{eq:spatial_frequencies}
	\varpi_{\xi_{i}}
	=
	\frac{2 \pi}{L}
	\times
	\begin{dcases}
		\xi_{i} + 1 / 2 & \text{for ABC} \, ,
		\\[0.25em]
		\xi_{i} & \text{for PBC} \, .
	\end{dcases}
\end{equation}

When implementing a finite spatial volume in practical DSE calculations, it is
\begin{itemize}
	\item[(i)] beneficial to rearrange the sum in Eq.~\eqref{eq:int_to_sum}
	such that it resembles a spherical coordinate system \cite{Fischer:2002eq};
	\item[(ii)] vital to remove finite-size artifacts, which are caused by the
	mismatch of the discrete momentum grid and the $\OO(3)$-symmetric continuum
	at large momenta;
	\item[(iii)] necessary to pay attention to the proper inclusion of the
	zero mode $\vec{q}_{\vec{\xi}=\vec{0}} = \vec{0}$ that appears for PBC.
\end{itemize}
Since a detailed discussion of (i)--(iii) can be found in our previous work
\cite{Bernhardt:2021iql}, we shall keep this section concise and refer the
reader to that work for more details.

\subsection{\label{framework:dse}%
	Dyson--Schwinger equations
}

Generally, the functional approach works with differential and/or integral
equations, relating loops of nonperturbative correlation functions with each
other. These can be formulated in configuration space or, more commonly,
in momentum space.

As in our previous work \cite{Bernhardt:2021iql}, we use DSEs to investigate
QCD in a finite, three-dimensional cubic volume and at nonzero temperature $T$
and quark chemical potential $\mu_{f}$, where $f$ labels the flavor. The central
quantity is the dressed quark propagator $S_{f}$ that takes the form%
\footnote{%
	We work within the Matsubara (imaginary time) formalism,
	i.e., in Euclidean space-time with positive metric signature (++++).
	The gamma matrices are Hermitian and obey
	$\{\gamma_\nu, \gamma_\sigma\} = 2 \+ \delta_{\nu\sigma}$.
}
\begin{align}
	\label{eq:quark_prop}
	S_{f}^{-1}(\omega_{n}, \vec{p}_{\vec{\zeta}})
	&=
	\ii \+ \gamma_{4} \+ (\omega_{n} + \ii \+ \mu_{f}) \+
	C_{f}(\omega_{n}, \vec{p}_{\vec{\zeta}})
	\notag
	\\
	&\phantom{=\;}
	+
	\ii \+ \vec{\gamma} \cdot \vec{p}_{\vec{\zeta}} \+
	A_{f}(\omega_{n}, \vec{p}_{\vec{\zeta}})
	+
	B_{f}(\omega_{n}, \vec{p}_{\vec{\zeta}})
\end{align}
with temporal Matsubara frequencies $\omega_{n} = (2 \+ n + 1) \+ \pi T$,
$n \in \bbZ$, and $\vec{p}_{\vec{\zeta}} = [\varpi_{\zeta_{1}},
\varpi_{\zeta_{2}}, \varpi_{\zeta_{3}}]^{\top}$, $\vec{\zeta} \in \bbZ^{3}$.
The dressing functions $A_{f}$, $B_{f}$, and $C_{f}$ carry the nonperturbative
information, which manifests in a nontrivial dependence on the Matsubara
frequency and three-momentum.

The dressed quark propagator is the solution of its corresponding DSE that
reads
\begin{align}
	\label{eq:quark_dse}
	S_{f}^{-1}(\omega_{n}, \vec{p}_{\vec{\zeta}})
	&=
	Z_{2} \+
	\bigl(
	\+ \ii \+ \gamma_{4} \+ (\omega_{n} + \ii \+ \mu_{f})
	+
	\ii \+ \vec{\gamma} \cdot \vec{p}_{\vec{\zeta}}
	+
	Z_{m} \+ m_{f}
	\bigr)
	\notag
	\\
	&\phantom{=\;}
	-
	\Sigma_{f}(\omega_{n}, \vec{p}_{\vec{\zeta}}) \, ,
\end{align}
where $Z_{2}$, $Z_{m}$, and $m_{f}$ denote the wave function renormalization
constant, mass renormalization constant, and current quark mass, respectively.
The former are determined in vacuum using a momentum-subtraction scheme. The
self-energy is explicitly given by
\begin{align}
	\label{eq:quark_self-energy}
	\Sigma_{f}(\omega_{n}, \vec{p}_{\vec{\zeta}})
	&=
	(\ii \+ \gs)^{2} \+
	\frac{4}{3}
	\frac{Z_{2}}{\tilde{Z}_{3}}
	\frac{T}{L^{3}}
	\sum_{k \in \bbZ}
	\sum_{\vec{\xi} \in \bbZ^{3}}
	D_{\nu\rho}(\omega_{k} - \omega_{n},
	\vec{q}_{\vec{\xi}} - \vec{p}_{\vec{\zeta}})
	\notag
	\\[0.25em]
	&\phantom{=\;}
	\times
	\gamma_{\nu} \+ S_{f}(\omega_{k}, \vec{q}_{\vec{\xi}}) \+
	\Gamma_{\rho}^{f}(\omega_{k}, \vec{q}_{\vec{\xi}},
	\omega_{n}, \vec{p}_{\vec{\zeta}})
\end{align}
with the strong coupling constant $\gs$, ghost renormalization constant
$\tilde{Z}_{3}$, and dressed gluon propagator $D_{\nu\rho}$; two factors of
$\ii\+\gs$  from the vertices appear explicitly such that $\Gamma_{\rho}^{f}$
denotes the reduced dressed quark-gluon vertex. Color d.o.f.\ are already
traced out resulting in the prefactor $4 / 3$ for $\Nc = 3$ colors.

\begin{figure}[t]
	\centering%
	\includegraphics[scale=1.0]{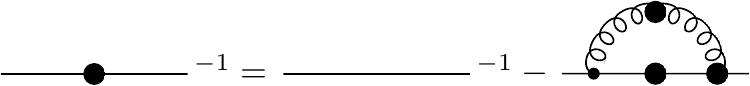}%
	\caption{\label{fig:quark_dse}%
		The DSE for the quark propagator. Large filled dots indicate dressed
		quantities; solid and curly lines represent quark and gluon propagators,
		respectively. There is a separate DSE for each flavor.
	}%
	\vspace*{1em}%
	\includegraphics[scale=1.0]{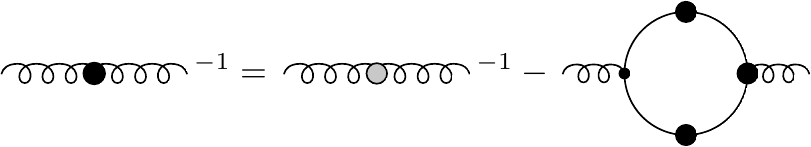}%
	\caption{\label{fig:gluon_dse}%
		Truncated gluon DSE. The large gray dot denotes the quenched gluon
		propagator that is taken from the lattice while the quark loop is
		evaluated explicitly. The latter contains an implicit flavor sum.
	}%
\end{figure}

Equation \eqref{eq:quark_dse} is shown diagrammatically
in Fig.~\ref{fig:quark_dse}. In order to solve this self-consistent equation
for the dressed quark propagator $S_{f}$, we need to specify two yet unknown
quantities: the dressed gluon propagator and the dressed quark-gluon vertex,
which both fulfill their own DSEs. The truncation used here has been studied
extensively in the past (see, e.g., Refs.~\cite{Isserstedt:2019pgx,%
Gunkel:2019xnh,Isserstedt:2020qll,Gunkel:2021oya,Bernhardt:2021iql} for recent
works and Ref.~\cite{Fischer:2018sdj} for a review) and is characterized as
follows.

First, in the full gluon DSE, we replace all pure Yang--Mills diagrams
by fits to quenched, temperature-dependent lattice results
\cite{Fischer:2010fx,Maas:2011ez} while the quark loop is evaluated explicitly,
thereby unquenching the gluon. The resulting equation is shown in
Fig.~\ref{fig:gluon_dse}. As a consequence, the quark and gluon DSEs are
nontrivially coupled. This results in a temperature and (implicit)
chemical-potential dependence of the gluon beyond modeling; moreover, the
gluon becomes sensitive to the chiral dynamics of the quarks. Though this
approximation misses implicit quark-loop effects in the Yang--Mills
self-energies, these are subleading in a skeleton expansion in comparison to
the quark loop. The impact of this approximate yet efficient way to compute
the unquenched gluon can be estimated in vacuum within the framework of
Ref.~\cite{Fischer:2003rp} and is found to be below the five-percent level.
Furthermore, finite-volume effects in the quenched gluon propagator have been
studied extensively at zero temperature both on the lattice and within DSEs.
In both frameworks, substantial volume effects have only been found at very
small momenta \cite{Fischer:2007pf,Bogolubsky:2007ud,Cucchieri:2007md}.
Since these are irrelevant for nonzero-temperature calculations due to the
additional temperature scale, we work with the hypothesis that results of the
present study will not be altered by the inclusion of volume effects in the
gluon input in future studies.

Second, for the dressed quark-gluon vertex, we use an ansatz that supplements
the leading term of the Ball--Chiu vertex \cite{Ball:1980ay} with a
phenomenological dressing function. The latter is constructed based on a
Slavnov--Taylor identity for the full vertex together with the requirement of
the correct logarithmic running of the propagators in the ultraviolet momentum
region. As a consequence, this construction also takes volume effects in the
vertex via the quark dressing functions imposed by the Slavnov--Taylor identity
into account.

For the sake of brevity and since our setup is identical to the one used
in previous works, we do not show explicit expressions and refer the reader
to Refs.~\cite{Fischer:2018sdj,Isserstedt:2019pgx,Bernhardt:2021iql} and
references therein for more details.%
\footnote{%
	The fits to the quenched lattice gluon propagator can be found in the
	earlier work \cite{Eichmann:2015kfa}, and the quark-gluon vertex ansatz
	is discussed in detail in Ref.~\cite{Fischer:2014ata}.
}

The resulting set of truncated DSEs for the quark and gluon propagators is
solved numerically. We use $2 + 1$ quark flavors and work in the
isospin-symmetric limit for the two light quarks, i.e., $\massu = \massd$
and $\muu = \mud$, and choose $\mus = 0$. Thus, the baryon chemical potential
is given by $\muB = 3 \+ \muu$. The current quark masses are
determined as follows (see Refs.~\cite{Fischer:2014ata,Isserstedt:2019pgx}
for more details): (i) the up/down quark mass is chosen such that the
high-temperature behavior of the regularized quark condensate matches lattice
results \cite{Borsanyi:2010bp}; (ii) the strange quark mass follows from the
ratio $\masss / \massu = 25.7$, which is obtained from in-vacuum pion and kaon
masses using the Bethe--Salpeter framework of Ref.~\cite{Heupel:2014ina}.

Finally, we would like to discuss the issue of the potential effects of
(space-like) correlations with the quantum numbers of mesons in the quark DSE.
Within the DSE framework, these correlations appear as part of a particular
diagram in the DSE for the quark-gluon vertex and therefore feed back into the
quark DSE. These correlations have been identified and explored in a number of
works in the vacuum \cite{Fischer:2007ze,Fischer:2008wy,Fischer:2008sp} as well
as at finite temperature and chemical potential
\cite{Fischer:2011pk,Gunkel:2021oya}. In Ref.~\cite{Gunkel:2021oya}, it was
found that the inclusion of meson backcoupling effects onto the quark has only
very little effect on the location of the CEP in the QCD phase diagram. On the
other hand, however, these degrees of freedom will develop long-range
correlations inside the critical regions around second-order transitions and
will therefore be crucial when it comes to the calculation of anomalous
dimensions and critical scaling.

Due to the substantial technical complications that arise when including these
contributions, we ignore those in the present exploratory work. As already
discussed in the introduction, we therefore do not expect to be able to study
all (and likely not even the most important) contributions to volume effects
inside the critical scaling region around the CEP. However, outside this region,
these contributions are known to be subleading. Thus, our results should be
meaningful and relevant---in particular since beyond-mean-field calculations
indicate that the  size of the critical region is rather small
\cite{Schaefer:2004en,Schaefer:2007pw}.

\subsection{\label{framework:fluctuations}%
	Quark and baryon number fluctuations
}

In the following, we first briefly summarize some general aspects of
fluctuations (see, e.g., Refs.~\cite{Asakawa:2015ybt,Luo:2017faz} for reviews)
and then detail how we determine these from our solutions of the DSEs specified
in the previous subsection.

In three-flavor QCD with quark chemical potentials $\muu$, $\mud$, and $\mus$,
the quark number fluctuations are (dimensionless) derivatives of QCD's
thermodynamic potential $\Omega$ with respect to these chemical potentials, viz.
\begin{equation}
	\label{eq:chi_q}
	\chi_{ijk}^{\textup{uds}}
	=
	-\frac{1}{T^{4 - (i + j + k)}}
	\frac{\partial^{i + j + k} \+ \Omega}
	{\partial \muu^{i} \+ \partial \mud^{j} \+ \partial \mus^{k}}
\end{equation}
with $\Omega = -T \log(\calZ) / \+ V$ and $i,\+ j,\+ k \in \bbN$, where $\calZ$
denotes the grand-canonical partition function of QCD and $V$ the volume of the
system. The quark chemical potentials are related to the ones for baryon number
(B), electric charge (Q), and strangeness (S) via
$\muu = \muB / 3 + 2 \+ \muQ / 3$, $\mud = \muB / 3 - \muQ / 3$, and
$\mus = \muB / 3 - \muQ / 3 - \muS$. With these relations, the corresponding
fluctuations $\chi_{ijk}^{\textup{BQS}}$ can be expressed as linear
combinations of quark number fluctuations. For example,
\begin{equation}
	\label{eq:chi_2_B}
	\chinX{2}{B}
	=
	\frac{1}{9} \,
	\bigl[
	\chinX{2}{u}
	+
	\chinX{2}{d}
	+
	\chinX{2}{s}
	+
	2 \, \bigl(
	\chinX{11}{ud}
	+
	\chinX{11}{us}
	+
	\chinX{11}{ds}
	\bigr)
	\bigr] \, .
\end{equation}

Generally, fluctuations of conserved charges are sensitive to phase transitions
and thus expected to be prime candidates to provide signatures of the
(conjectured) CEP in experiments. Ratios are particularly interesting because
they are equal to ratios of cumulants of the corresponding probability
distributions that can be extracted from heavy-ion collisions by means of
event-by-event analyses; see Refs.~\cite{Koch:2008ia,Asakawa:2015ybt,%
Luo:2017faz,Bzdak:2019pkr} and references therein for more details.
Prominent ratios are the ones involving the skewness and kurtosis, namely
\begin{equation}
	\label{eq:chi_ratios}
	\frac{\chinX{3}{B}}{\chinX{2}{B}}
	=
	S_{\textup{B}} \+ \sigma_{\textup{B}}
	\, , \qquad
	\frac{\chinX{4}{B}}{\chinX{2}{B}}
	=
	\kappa_{\textup{B}} \+ \sigma_{\textup{B}}^{2}
	\, ,
\end{equation}
where $\sigma_{\textup{B}}^{2}$, $S_{\textup{B}}$, and $\kappa_{\textup{B}}$
denote the variance, skewness, and kurtosis of the net-baryon distribution,
respectively. Analogous expressions hold for charge and strangeness.

In this work, we consider $\chinX{2}{B}$ and the skewness and kurtosis
ratios, which are determined via the quark number fluctuations
[see Eq.~\eqref{eq:chi_2_B}]. As in our previous work
\cite{Isserstedt:2019pgx}, we compute the latter from the quark number
densities $n_f = -\partial \+ \Omega \+ / \partial \mu_{f}$; for instance,
\begin{equation}
	\label{eq:chi_from_n}
	\chinX{2}{u}
	=
	\frac{1}{T^2}
	\frac{\partial n_{\textup{u}}}{\partial \muu} \, .
\end{equation}

In infinite volume, the regularized quark number density for a flavor $f$ is
obtained from the dressed quark propagator via \cite{Isserstedt:2019pgx}
\begin{equation}
	\label{eq:n_f_reg}
	n_{f}^{\textup{reg}}
	=
	-\Nc \+ Z_{2} \+ K_{f}
\end{equation}
with
\begin{align}
	\label{eq:n_f_reg_kernel_1}
	K_{f}
	&=
	T \sum_{k \in \bbZ} \,
	\int \frac{\dd^{3} q}{(2 \pi)^{3}}
	\Tr\bigl[ \gamma_{4} \+ S_{f}(\omega_{k}, \vec{q}) \bigr]
	\notag
	\\[.25em]
	&\phantom{=\;}
	-
	\int_{-\infty}^{\infty} \frac{\dd q_{4}}{2 \pi} \,
	\int \frac{\dd^{3} q}{(2 \pi)^{3}}
	\Tr\bigl[ \gamma_{4} \+ S_{f}(q_{4}, \vec{q}) \bigr]
	\notag
	\\[.25em]
	&\equiv
	\Ksum - \Kint \, .
\end{align}
The term $\Ksum$ contains a sum over the temporal Matsubara frequencies and
is the one expected to yield the quark number density. However, it is divergent
and needs to be regularized if---as in our approach---evaluated numerically
where only a finite number of Matsubara frequencies is available. To this end,
we subtract the term $\Kint$ that does not depend explicitly on temperature or
chemical potential and is therefore known as a ``vacuum contribution''
\cite{Kapusta:2006pm}. This subtraction scheme is based on the Euclidean
version of the contour-integration technique for Matsubara sums; see, e.g.,
Refs.~\cite{Gao:2016hks,Isserstedt:2019pgx,Isserstedt:2021acw} for more details.

\begin{figure*}[t]
	\centering%
	\includegraphics[scale=1.0]{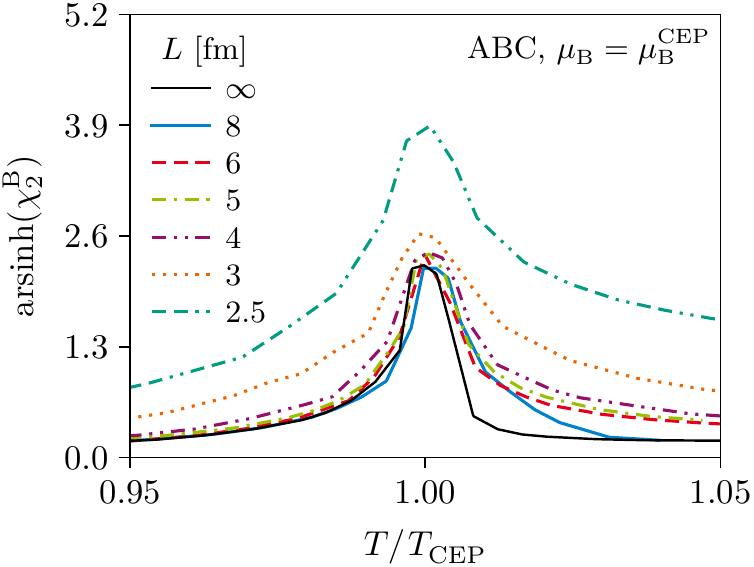}%
	\hspace*{1em}%
	\includegraphics[scale=1.0]{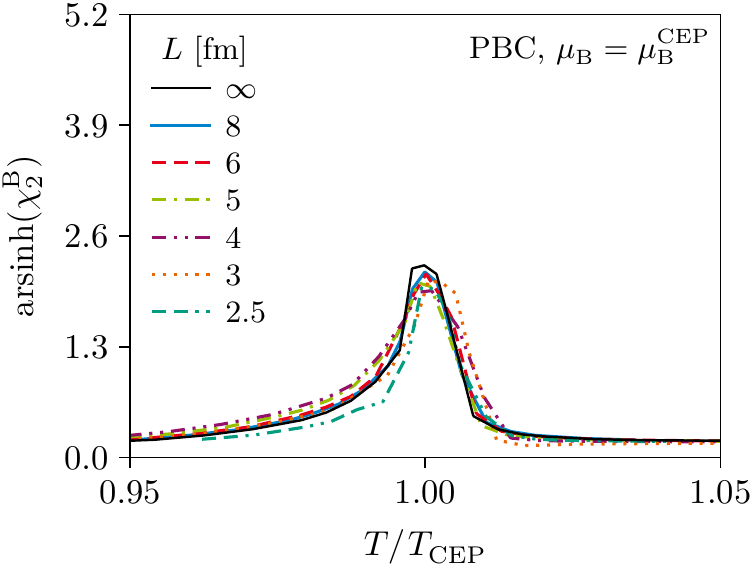}
	\caption{\label{fig:chi2}%
		Volume dependence of the second-order baryon number fluctuation
		for ABC (left) and PBC (right) in the vicinity of the CEP. We plot the
		fluctuations at the volume-dependent chemical potential of the CEP,
		$\muB = \muCEP(L)$, as functions of the reduced temperature
		$T / \+ \TCEP$, where $\TCEP = \TCEP(L)$ is the volume-dependent
		temperature of the CEP. For better visibility, we display the inverse
		hyperbolic sine ($\arsinh$) of the fluctuations, which resembles
		a logarithmic plot in both positive and negative direction.
	}%
\end{figure*}

In a finite volume, the expression for $K_f$ turns into
\begin{align}
	\label{eq:n_f_reg_kernel_2}
	K_{f}
	&=
	\frac{T}{L^{3}}
	\sum_{k \in \bbZ}
	\sum_{\vec{\xi} \in \bbZ^{3}}
	\Tr\bigl[ \gamma_{4} \+ S_{f}(\omega_{k}, \vec{q}_{\vec{\xi}}) \bigr]
	\notag
	\\[.25em]
	& \phantom{=\;}
	-
	\int_{-\infty}^{\infty} \frac{\dd q_{4}}{2 \pi} \,
	\int_{\textup{ABC\+/\+PBC}} \frac{\dd^{3} q}{(2 \pi)^{3}}
	\Tr\bigl[ \gamma_{4} \+ S_{f}(q_{4}, \vec{q}) \bigr] \, ,
\end{align}
where we replaced the infinite-volume integral by the finite-volume sum
with respect to spatial Matsubara sums in the first term $\Ksum$ but not in
the subtraction term $\Kint$. It turns out that this procedure is necessary to
avoid artifacts in the resulting densities related to the high-momentum
behavior of the subtraction term. The subscript ABC/PBC of the integral
indicates that we treat the radial integral differently for different boundary
conditions: whereas there is no momentum gap for PBC and the integral
consequently starts at zero, for ABC we have
$\vec{q}_{\vec{\xi} = \vec{0}}^{\textup{ABC}} \neq \vec{0}$ and therefore set
the lower integration limit to the value of the smallest possible momentum
magnitude.

Before presenting our results, we would also like to briefly comment on
numerical stability. First, due to the necessary subtraction, the quark
number density itself is numerically rather sensitive already in the
infinite-volume setup and, as it turns out, even more so in finite-volume
calculations. Second, its derivatives, the fluctuations, are at present only
accessible by means of finite-difference formulae. This leads to a limited
accuracy with numerical uncertainties of the order of ten percent.

\section{\label{results}%
	Results and discussion
}

Our starting point is the infinite-volume result for the location of the CEP
from Ref.~\cite{Isserstedt:2019pgx}, which is displayed in the upper diagram
of Fig.~\ref{fig:phase_diagrams} together with recent results from other
functional approaches \cite{Fu:2019hdw,Gao:2020qsj,Gao:2020fbl,Gunkel:2021oya}
and an extrapolation obtained from lattice QCD \cite{Bellwied:2015rza}. As
already discussed in the introduction, all results from these state-of-the-art
functional calculations cluster in the region
$(\muCEP,\, \TCEP) = (\numrange{495}{654}, \numrange{108}{119}) \, \si{\MeV}$.
The size of this region can be viewed as a measure of the lower bound for the
remaining systematic error of the functional approach because the employed
truncations differ in many technical details but are roughly on a similar level
regarding the overall treatment of the tower of FRG equations and DSEs.

In the following, we discuss finite-volume baryon number fluctuations and their
ratios around the CEP and compare them to their infinite-volume limit, which is
explicitly calculated using the framework of Ref.~\cite{Isserstedt:2019pgx}
(black dot in Fig.~\ref{fig:phase_diagrams}). We study cubes with edge lengths
of $L = 2.5$, $3$, $4$, $5$, $6$, and $\SI{8}{\fm}$ for both ABC and PBC.

We recall that the location of the CEP in the phase diagram is volume
dependent---the corresponding calculations have been carried out in
Ref.~\cite{Bernhardt:2021iql} and are shown again in the lower diagrams of
Fig.~\ref{fig:phase_diagrams}. Sizable effects only occur for volumes
$V \lesssim (\SI{5}{\fm})^3$ and are much larger for ABC than for PBC. For the
sake of comparability, in the following we always show results obtained around
the respective critical chemical potential for each system size. Additionally,
we also normalize all temperatures to the corresponding critical temperatures.
Since fluctuations around the CEP vary rapidly with temperature, a dense
numerical grid is necessary to avoid misalignments. We accomplish this by
using steps of one MeV in temperature.

In the interpretation of all results in this section, the reader should
keep in  mind that the present calculation is exploratory because effects from
potential long-range and critical correlations (such as the sigma meson) are,
as detailed in the previous section, not yet taken into account.

\subsection{\label{results:fluctuations}%
	Baryon number fluctuations
}

\begin{figure*}[t]
	\centering%
	\includegraphics[scale=1.0]{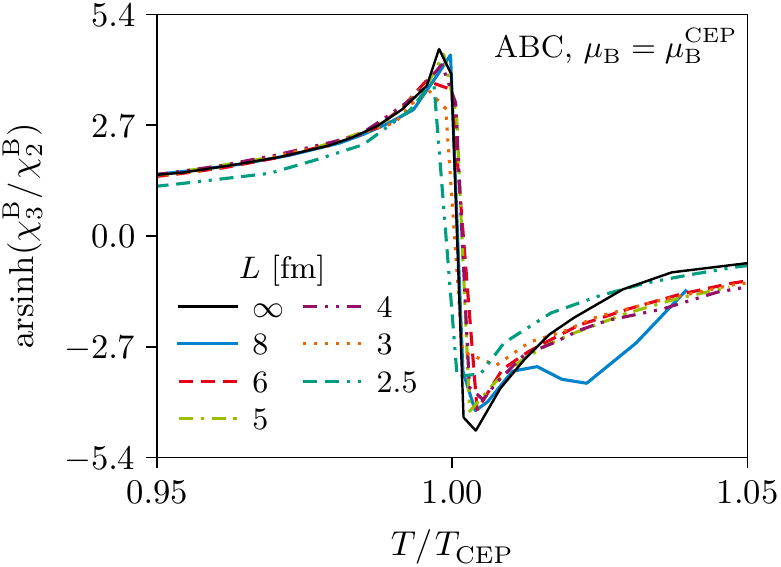}%
	\hspace*{1em}
	\includegraphics[scale=1.0]{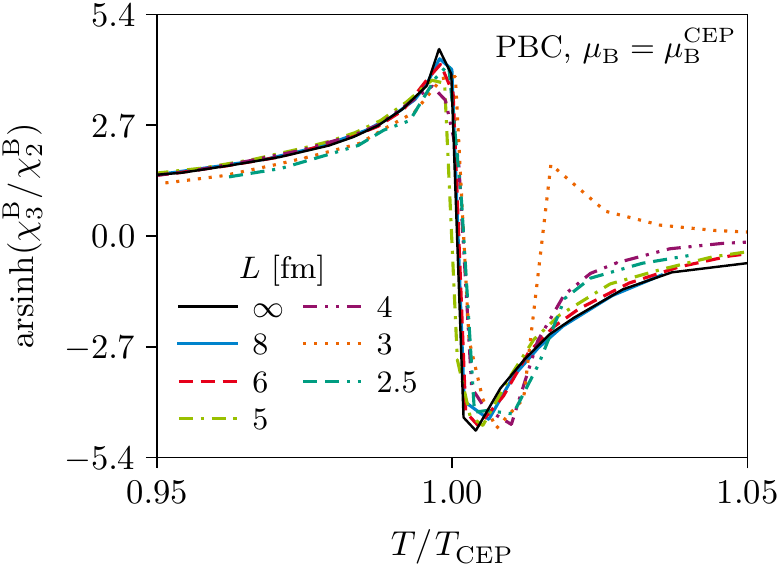}%
	\\[1em]%
	\includegraphics[scale=1.0]{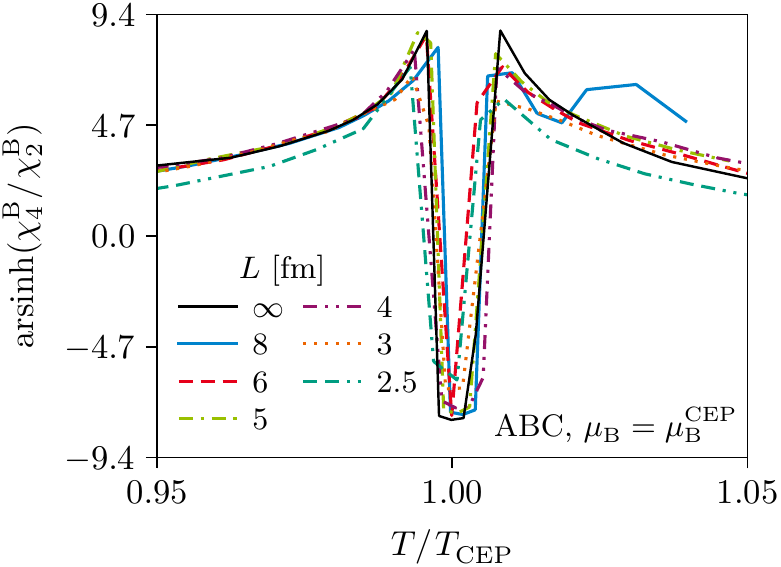}%
	\hspace*{1em}%
	\includegraphics[scale=1.0]{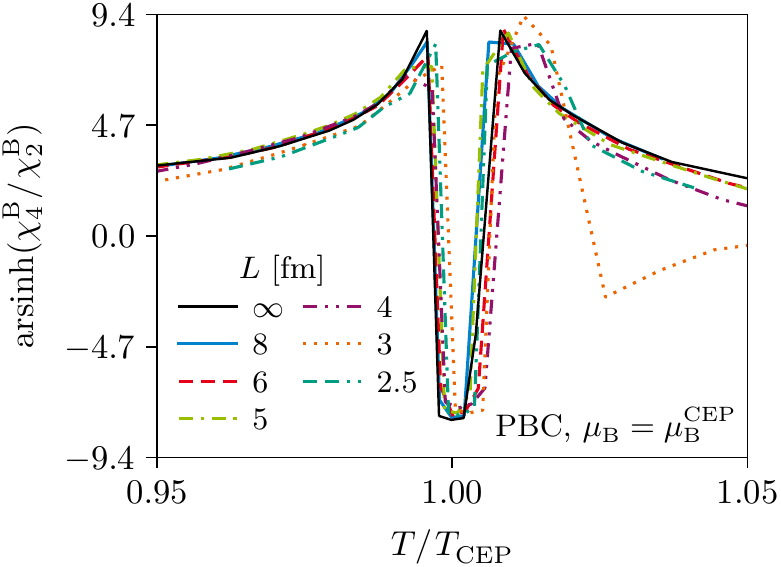}%
	\caption{\label{fig:ratios}%
		Volume dependence of the skewness (top) and kurtosis (bottom) ratios
		for ABC (left) and PBC (right) in the vicinity of the CEP. The way
		we plot the fluctuations is identical to Fig.~\ref{fig:chi2}.
	}%
\end{figure*}

We begin our discussion with the results for the baryon number fluctuations.
In Fig.~\ref{fig:chi2}, we exemplarily show the second-order baryon number
fluctuation $\chinX{2}{B}$ for both ABC (left) and PBC (right) at finite
volume as well as the infinite-volume result in black for comparison.

Before we discuss the details, a general comment is in order. As visible
in both plots, the infinite-volume result does not show the expected divergence
at the critical temperature. This is entirely due to our limited resolution
in temperature. Given an unlimited amount of CPU time, we could determine the
location of the CEP with arbitrary precision and perform calculations
arbitrarily close to the CEP, thereby extracting the point of divergence
exactly. In practice, for now, this is the best we could do and thus the
finite width and height of the peak at infinite volume serves as our control
quantity for the size of potential effects at finite volume.

Starting with the ABC results, we find clearly visible volume effects. First,
we see a monotonous increase of $\chinX{2}{B}$ with decreasing system size
across the whole temperature range. Second, we find that the ratio of the peak
height to the tails of the peaks, e.g., $\chinX{2}{B}(1.00) / \chinX{2}{B}(1.05)$,
grows continuously with increasing system size (even within the resolution limits
discussed above). The $L = \SI{2.5}{\fm}$ line especially stands out with the
$\arsinh$ of its peak value nearly being doubled as compared to the infinite-volume
result---its peak-to-tail ratio, however, is much smaller. For $T < \TCEP$, we find
a consistent infinite-volume limit, i.e., the results for $L \gtrsim \SI{5}{\fm}$
are very similar to one another, while the $L = \SI{8}{\fm}$ line coincides
with the infinite-volume one. For $T > \TCEP$, however, we see noticeable
deviations between the finite- and infinite-volume results. We have not
succeeded to track  down these deviations unambiguously, but we believe they
originate from the finite-volume adjusted subtraction procedure of the density
outlined in Sec.~\ref{framework:fluctuations} either as remnants or as an
overcompensation of the initial problem.

In contrast, the fluctuations using PBC shown in the right diagram of
Fig.~\ref{fig:chi2} are much less dependent on the volume of the system than
the fluctuations with ABC. This resembles similar differences between PBC and
ABC effects on the location of the CEP discussed above. In fact, the volume
dependence of the fluctuations with PBC are within our margin of error.
Nonetheless, for $T < \TCEP$, one can observe a monotonous increase of
$\chinX{2}{B}$ with decreasing system size down to $L = \SI{4}{\fm}$. For
smaller box sizes, the values start decreasing again. This behavior is similar
to the nonmonotonous volume dependence of the location of the CEP around
$L = \SI{3}{\fm}$ that has been already noticed in
Ref.~\cite{Bernhardt:2021iql}. It may be linked to the onset of the
so-called epsilon regime at very small volumes \cite{Leutwyler:1992yt}.

In addition, the infinite-volume limit of the PBC fluctuations is much more
consistent for $T > \TCEP$ as compared to ABC. That is, there are no
substantial deviations between the $L = \SI{8}{\fm}$ and the $L \to \infty$
lines. As a consequence, one might conjecture that the numerical problems of
ABC for $T > \TCEP$ are connected to infrared momentum modes because ABC
introduce an effective infrared cutoff,
$\vec{q}_{\vec{\xi} = \vec{0}}^{\textup{ABC}} \neq \vec{0}$,
while PBC do not, $\vec{q}_{\vec{\xi} = \vec{0}}^{\textup{PBC}} = \vec{0}$.

We also extracted the peak heights of the ABC and PBC results for
$\chinX{2}{B}$ as a function of volume and analyzed their behavior in terms
of power laws. Whereas the PBC results are virtually independent of volume,
i.e., roughly proportional to $V^0$, there is no clear power-law behavior
visible for the ABC results if all volumes are considered. This may or may not
be connected to the numerical error discussed at the end of
Sec.~\ref{framework:fluctuations} and needs to be revisited in the future
in a more refined framework. Interestingly, as we will confirm in the next
subsection, the volume dependence is similar for the higher-order fluctuations,
too.

\subsection{\label{results:ratios}%
	Ratios of baryon number fluctuations
}

Next, we turn to ratios of baryon number fluctuation. To this end, we display
(the inverse hyperbolic sine of) both the skewness and kurtosis ratios for
ABC and PBC in Fig.~\ref{fig:ratios}. For comparison, the infinite-volume
result is again shown as a black line. In general, one can observe that both
ratios for both boundary conditions and all system sizes qualitatively coincide
very well for $T \leq \TCEP$. Additionally, they are also compatible with the
respective infinite-volume results. Furthermore, the sign changes in both
ratios are consistent with predictions based upon general grounds
\cite{Asakawa:2009aj,Stephanov:2011pb}.

For ABC, however, we find again slight inconsistencies between the finite- and
infinite-volume data for $T > \TCEP$. Neglecting the obvious outlier at
$L = \SI{8}{\fm}$, the lines of the larger volumes deviate qualitatively from
the $L \to \infty$ one, especially for larger temperatures. This deviation is
more pronounced for the skewness ratio. Contrary to this, the PBC
results exhibit once more no such behavior, and we observe a consistent
infinite-volume limit also for $T > \TCEP$. In fact, the $L = \SI{8}{\fm}$ and
the infinite-volume curves are almost indistinguishable. This seems to
corroborate our assumption that the infrared cutoff of ABC leads to some
numerical problems for $T > \TCEP$.

In addition to that, there are two notable outliers: the curves for ABC at
$L = \SI{8}{\fm}$ and PBC at $L = \SI{3}{\fm}$. Due to the randomness in their
occurrence, they are most likely of purely numerical origin. We remark that the
deviation from the rest of the curves in both cases occurs again for
$T > \TCEP$, which makes a connection to the subtraction procedure plausible.
This also implies that PBC are not completely immune to these numerical
problems.

Overall, we find the remarkable result that all individual volume dependences
of the fluctuations cancel once ratios are studied. This not only true for
large volumes but also for our smallest system sizes of $L=5$, $4$, $3$, and
$\SI{2.5}{\fm}$. This is somewhat in contrast to the results of
Ref.~\cite{Almasi:2016zqf}, where significant volume effects in the kurtosis
ratio have been found for volumes $V \lesssim (\SI{5}{\fm})^3$ within an FRG
treatment of the quark-meson model. Since the two approaches are rather
different, e.g., our approach treats the gluonic sector explicitly but neglects
a class of mesonic fluctuations (as discussed at the end of
Sec.~\ref{framework:dse}), and vice versa in their approach, it may be
interesting to provide a systematic comparison in the future.

\section{\label{summary}%
	Summary and conclusions
}

In this work, we determined the volume dependence of the skewness and kurtosis
ratios of baryon number fluctuations within a well-established functional
approach to QCD. For a wide range of cubic spatial volumes with edge lengths
between $L = \SI{2.5}{\fm}$ and $L = \SI{8}{\fm}$ and two different boundary
conditions (ABC and PBC), we observe almost no volume dependence of these
ratios. This is a highly nontrivial result because the individual results for
the different fluctuations, both for ABC and for PBC, reveal a pattern that is
at odds with the general expectation of linear dependence on volume:
whereas the PBC results for $\chinX{2}{B}$ do not change with volume, the ones
for ABC are even inversely proportional to $V = L^3$. Nevertheless, all these
dependences cancel in the ratios, which are important when comparing with
experimental results from heavy-ion collisions.

As explained in the main part of this exploratory work, our results need
to be put into perspective. Mesonic degrees of freedom that are crucial to
obtain the correct critical scaling inside the critical region around the CEP
are not yet included. However, they have been taken into account in the
infinite-volume calculation of the location of the CEP of
Ref.~\cite{Gunkel:2021oya}. Thus, their inclusion in our finite-volume
study is in principle possible, although very demanding in terms of CPU time.
This is left for future work.

\section*{\label{acknowledgments}%
	Acknowledgments
}

We thank Jana N.~Guenther for enlightening discussions. This work has been
supported by the Helmholtz Graduate School for Hadron and Ion Research
(HGS-HIRe) for FAIR, the GSI Helmholtzzentrum f\"{u}r Schwerionenforschung,
and the Bundesministerium f\"{u}r Bildung und Forschung (BMBF) under Contract
No.~05P18RGFCA. Feynman diagrams were drawn with \textsc{JaxoDraw}
\cite{Binosi:2008ig}.

\bibliographystyle{elsarticle-num}
\bibliography{FiniteVolumeFluctuationsBibliography}

\end{document}